\documentclass[12pt]{article}
\usepackage[margin=1in]{geometry}
\usepackage{setspace}
\usepackage{hyperref}
\usepackage{parskip}

\title{OpCode-Based Malware Classification Using Machine Learning and Deep Learning Techniques}
\author{Varij Saini, Rudraksh Gupta, Neel Soni \\ Students, Cybersecurity and Threat Intelligence \\ University of Guelph}
\date{April 1, 2025}

\begin{document}
\maketitle

\section*{Executive Summary}
This technical report presents a comprehensive analysis of malware classification using OpCode sequences. Two distinct approaches are evaluated: traditional machine learning using n-gram analysis with Support Vector Machine (SVM), K-Nearest Neighbors (KNN), and Decision Tree classifiers; and a deep learning approach employing a Convolutional Neural Network (CNN). The traditional machine learning approach establishes a baseline using handcrafted 1-gram and 2-gram features from disassembled malware samples. The deep learning methodology builds upon the work proposed in ``Deep Android Malware Detection'' by McLaughlin et al. and evaluates the performance of a CNN model trained to automatically extract features from raw OpCode data. Empirical results are compared using standard performance metrics (accuracy, precision, recall, and F1-score). While the SVM classifier outperforms other traditional techniques, the CNN model demonstrates competitive performance with the added benefit of automated feature extraction.

\section{Introduction}
In an era of increasingly sophisticated cyberattacks, the detection and classification of malicious software, or malware, remain central to safeguarding digital infrastructure. Among the most formidable threats are Advanced Persistent Threats (APTs), which employ stealthy and prolonged campaigns to exfiltrate sensitive data and compromise critical systems \cite{67, 1h}. Traditional signature-based antivirus mechanisms have shown declining efficacy against such advanced adversaries, as malware authors continually develop obfuscation techniques to evade static and dynamic detection engines. In response, the cybersecurity research community has increasingly turned to machine learning (ML) and deep learning (DL) techniques, aiming to detect malware by uncovering subtle, intrinsic patterns within executable code \cite{1, 2}.

One promising line of inquiry involves the analysis of OpCode sequences—low-level machine instructions executed by a CPU during program execution. OpCodes offer a more granular and robust representation of a program’s behavior than higher-level features, making them less susceptible to evasion through superficial code manipulation. This project builds upon this insight, proposing a comparative evaluation of traditional machine learning algorithms and a deep learning-based model for malware classification based solely on OpCode sequences extracted from disassembled malware binaries \cite{3h,x1}.

The primary goal of this research is to assess the trade-offs between handcrafted n-gram features and automated feature learning. The traditional approach employs Support Vector Machines (SVM), K-Nearest Neighbors (KNN), and Decision Tree classifiers using 1-gram and 2-gram frequency distributions as input features. These models serve as a performance benchmark for the proposed deep learning method, which utilizes a Convolutional Neural Network (CNN) architecture to directly process raw OpCode sequences. By eliminating manual feature engineering, the CNN approach aims to streamline the malware detection pipeline while uncovering latent sequence patterns that may not be captured by fixed-length n-grams \cite{4,x2}.

To ensure methodological rigor, the models are trained and evaluated using standard classification metrics—accuracy, precision, recall, and F1-score—on a labeled dataset of APT malware families. Through this comparative study, we aim to provide empirical insights into the effectiveness of OpCode-based features, the scalability of automated learning in malware classification, and the practical implications for developing real-world threat detection systems. The research not only highlights the strengths and limitations of each approach but also establishes a reproducible framework for future work in OpCode-based malware analysis\cite{4h, 5}.
\subsection{Background and Motivation}
Malware continues to pose a critical threat to organizations worldwide, with evasion techniques rapidly evolving to bypass signature-based detection systems \cite{2h}. Static analysis of executable code structures, particularly through OpCode sequences, has emerged as a robust method for malware family identification. OpCodes, the machine-level instructions executed by processors, provide distinctive patterns even in obfuscated malware samples. The extraction and classification of these sequences form the foundation for effective threat intelligence and risk assessment. This project is designed to document Advanced Persistent Threat (APT) Tactics, Techniques, and Procedures (TTPs) and develop both traditional machine learning and deep learning models for detecting malicious payloads used by APT groups\cite{x3,7}.

\subsection{Research Objectives}
The study aims to:
\begin{itemize}
    \item Evaluate the effectiveness of OpCode n-gram analysis for malware classification.
    \item Compare the performance of traditional machine learning algorithms (SVM, KNN, Decision Tree) on 1-gram and 2-gram features.
    \item Implement and evaluate a CNN-based model for OpCode-based malware classification.
    \item Compare the performance of the CNN model with traditional classifiers to assess the trade-offs between manual feature engineering and automated deep learning feature extraction.
    \item Develop a reproducible methodology for OpCode-based malware classification.
\end{itemize}

\subsection{Scope and Limitations}
The research focuses on the static analysis of OpCode sequences extracted from disassembled malware samples. Dynamic analysis is beyond this study's scope \cite{6,x4}. While the evaluation is limited to three traditional classifiers and a CNN, further advanced models may yield improvements. The work provides empirical evidence and establishes a methodological framework for future malware analysis\cite{8,9}.

\section{Methodology}
This study employs both traditional machine learning and deep learning techniques to classify malware based on OpCode sequences extracted from disassembled APT malware samples. The dataset comprises files labeled by malware family, from which raw OpCodes are extracted while discarding operands. For traditional models, 1-gram and 2-gram frequency features are computed and transformed into normalized feature vectors. Class imbalance is addressed using RandomOverSampler. Support Vector Machine (SVM), K-Nearest Neighbors (KNN), and Decision Tree classifiers are trained and evaluated using standard metrics\cite{10, 11}.

In parallel, a Convolutional Neural Network (CNN) is implemented to process raw OpCode sequences without manual feature engineering. The CNN architecture includes two 1D convolutional layers, max pooling, ReLU activations, dropout regularization, and fully connected layers for classification. The model is trained using PyTorch with Adam optimization and learning rate scheduling. Performance of both approaches is evaluated using accuracy, precision, recall, and F1-score to assess effectiveness and generalization.
\subsection{Dataset Acquisition and Preprocessing}
The dataset comprises disassembled OpCode files extracted from malware samples belonging to various APT families. The files are organized by malware family using filenames formatted as \texttt{[malware\_family]\_[sample\_id].opcode}, enabling automated labeling during preprocessing.

\textbf{Traditional Machine Learning Approach}
\begin{itemize}
    \item \textbf{OpCode Extraction:} Files are parsed to extract OpCode sequences, ignoring operands.
    \item \textbf{N-gram Generation:} Both 1-gram (individual OpCodes) and 2-gram (sequential pairs) features are generated.
    \item \textbf{Feature Vector Creation:} Frequency distributions of n-grams are used to create feature vectors.
    \item \textbf{Data Normalization and Class Imbalance Handling:} StandardScaler is applied to normalize features, and RandomOverSampler is used to balance class distributions.
\end{itemize}

\textbf{Deep Learning Approach}
\begin{itemize}
    \item Preprocessed OpCode sequences are loaded from previously generated numpy arrays.
    \item Label encoding converts string labels to integers, and data is converted into PyTorch tensors with DataLoader objects created for batch processing.
\end{itemize}

\subsection{Feature Extraction and Representation}

Feature extraction in this study is tailored to the nature of OpCode sequences derived from disassembled malware binaries. For the traditional machine learning approach, features are created by generating n-grams—specifically 1-grams (individual OpCodes) and 2-grams (consecutive OpCode pairs). These n-grams are counted to form frequency distributions, which are then transformed into numerical feature vectors representing each malware sample. The resulting vectors are normalized using StandardScaler to ensure consistent scaling across features \cite{x5,12}.

In contrast, the deep learning approach eliminates the need for manual feature engineering. Raw OpCode sequences are preprocessed and converted into numerical arrays, which are then encoded into integer labels and transformed into PyTorch tensors. The CNN model directly processes these sequences, learning hierarchical features through convolutional layers. This method captures complex spatial patterns within OpCode sequences, enabling automated extraction of semantic features critical for malware classification. Both representations aim to preserve structural characteristics essential for accurate detection.

\textbf{Traditional Approach:}
\begin{verbatim}
def generate_ngrams(opcodes, n):
    return [" ".join(opcodes[i:i+n]) for i in range(len(opcodes) - n + 1)]
\end{verbatim}
Both unigram and bigram frequency counts are computed and merged for feature representation.

\textbf{Deep Learning Approach:}
Unlike the explicit n-gram generation used previously, the CNN model processes raw OpCode sequence data directly, eliminating the need for manual feature engineering.

\subsection{Classification Models}
This research explores two categories of classification models: traditional machine learning algorithms and a deep learning-based Convolutional Neural Network (CNN). In the traditional pipeline, three classifiers are implemented: Support Vector Machine (SVM) with a linear kernel and regularization parameter C=1, K-Nearest Neighbors (KNN) with 
k=3, and a Decision Tree limited to a maximum depth of 20. Additionally, a Voting Classifier ensemble is evaluated to assess the benefit of model aggregation.

The deep learning model is a 1D CNN architecture inspired by prior work in Android malware detection. It comprises two convolutional layers with kernel size 5, followed by max pooling, ReLU activations, and a dropout layer with a rate of 0.3 to mitigate overfitting. These layers feed into fully connected layers that output class probabilities. The CNN is trained using PyTorch with the Adam optimizer and dynamic learning rate scheduling to refine performance over multiple epochs.
\textbf{2.3.1 Traditional Machine Learning Models}\\
Three classifiers are implemented:
\begin{itemize}
    \item \textbf{SVM:} Linear SVM with \(C=1\).
    \item \textbf{KNN:} Implemented with \(k=3\).
    \item \textbf{Decision Tree:} A Decision Tree with a maximum depth of 20.
\end{itemize}
Additionally, a hard Voting Classifier ensemble is evaluated.

\textbf{2.3.2 Convolutional Neural Network (CNN) Model}\\
The CNN architecture follows the methodology of McLaughlin et al.:
\begin{itemize}
    \item \textbf{Convolutional Layers:} Two 1D convolutional layers with kernel size 5, capturing local patterns.
    \item \textbf{Max Pooling:} Reduces dimensionality and allows representation of variable-length OpCode sequences.
    \item \textbf{Activation and Dropout:} ReLU activations and a dropout rate of 0.3 to mitigate overfitting.
    \item \textbf{Fully Connected Layers:} Transition from convolutional features to classification outputs matching the number of malware families.
\end{itemize}

\textbf{CNN Model Architecture (Excerpt):}
\begin{verbatim}
class MalwareCNN(nn.Module):
    def __init__(self, input_dim, num_classes):
        super(MalwareCNN, self).__init__()
        self.conv1 = nn.Conv1d(in_channels=1, out_channels=64, kernel_size=5, stride=1, padding=2)
        self.conv2 = nn.Conv1d(in_channels=64, out_channels=128, kernel_size=5, stride=1, padding=2)
        self.maxpool = nn.MaxPool1d(kernel_size=2, stride=2)
        self.relu = nn.ReLU()
        self.dropout = nn.Dropout(0.3)
        with torch.no_grad():
            sample_input = torch.randn(1, 1, input_dim)
            sample_output = self._forward_features(sample_input)
            self.fc1_input_dim = sample_output.shape[1]
        self.fc1 = nn.Linear(self.fc1_input_dim, 128)
        self.fc2 = nn.Linear(128, num_classes)
    def _forward_features(self, x):
        x = self.relu(self.conv1(x))
        x = self.maxpool(x)
        x = self.relu(self.conv2(x))
        x = self.maxpool(x)
        x = torch.flatten(x, start_dim=1)
        return x
    def forward(self, x):
        x = x.unsqueeze(1)
        x = self._forward_features(x)
        x = self.relu(self.fc1(x))
        x = self.dropout(x)
        x = self.fc2(x)
        return x
\end{verbatim}

\subsection{Training and Evaluation}

\textbf{Traditional Approach:}\\
Models are trained on 80\% of resampled and scaled feature data using stratified sampling. Evaluation is conducted using accuracy, precision, recall, F1-score, and confusion matrices.\\
\textbf{Results Summary}
\begin{itemize}
    \item SVM: Accuracy 66.37\%, F1-score 64.04\%
    \item KNN: Accuracy 63.65\%, F1-score 61.02\%
    \item Decision Tree: Accuracy 62.14\%, F1-score 60.06\%
    \item Voting Classifier: Accuracy 68.61\% (but did not outperform SVM).
\end{itemize}

\textbf{Deep Learning Approach:}\\
The CNN is trained using PyTorch for 10 epochs with the Adam optimizer (lr=0.001) and a ReduceLROnPlateau scheduler. Evaluation metrics for the CNN are:
\begin{itemize}
    \item Accuracy: 62.14\%
    \item Precision: 64.49\%
    \item Recall: 62.14\%
    \item F1-score: 60.44\%
\end{itemize}
Comparative analysis indicates that while the CNN reduces the need for manual feature engineering, SVM remains the top-performing model on this dataset.

\section{Results and Analysis}

\subsection{Traditional Machine Learning Results}
Performance metrics are tabulated for SVM, KNN, Decision Tree, and the Voting Classifier. Confusion matrix analysis reveals that SVM has the most balanced performance across malware families.

\subsection{CNN Performance}
The CNN model shows rapid initial convergence with diminishing improvements after the fifth epoch. Final performance registers an accuracy of 62.14\%, with precision, recall, and F1-scores closely matching the Decision Tree classifier.

\section{Discussion}

\subsection{Comparative Analysis}
\begin{itemize}
    \item \textbf{Traditional vs. CNN:} SVM outperforms other traditional models and remains superior in overall classification performance. The CNN model offers benefits in automated feature learning but is limited on smaller datasets.
    \item \textbf{Feature Engineering Trade-Offs:} Manual n-gram generation captures detailed sequence relationships, while CNN eliminates this need but may require larger datasets and further architectural refinement.
\end{itemize}

\subsection{Practical Implications}
\begin{itemize}
    \item \textbf{Threat Intelligence Applications:} The ability to classify malware into distinct APT families aids in precise attribution and response.
    \item \textbf{Operational Integration:} Both approaches offer insights into building robust, automated detection systems, with SVM serving as a strong baseline and deep learning offering potential for future scaling.
\end{itemize}

\section{Future Road Map}
Building on the foundational work of OpCode-based malware classification, this project lays the groundwork for several promising research and development directions aimed at enhancing threat detection capabilities. As cyber threats continue to evolve, particularly from sophisticated Advanced Persistent Threat (APT) groups, it is imperative to develop adaptive, scalable, and intelligent systems that can offer real-time analysis and high-fidelity classification. The following roadmap outlines future milestones and key enhancements envisioned to improve the current methodology and expand its practical applicability \cite{x5}.

1. Integration of Dynamic and Hybrid Analysis:
While this project focused solely on static OpCode sequences, incorporating dynamic features—such as system call traces, network behavior, and runtime memory usage—can provide richer context and improve classification accuracy. A hybrid model that fuses static OpCode features with dynamic behavioral indicators could enable more robust malware detection, particularly for evasive or polymorphic threats.

2. Expansion of Dataset and Malware Families:
The effectiveness of machine learning models is often tied to the diversity and size of the training dataset. In future iterations, the dataset will be expanded to include a broader range of malware samples from public repositories (e.g., VirusShare, VirusTotal) and threat intelligence platforms. Incorporating additional malware families, especially those from emerging APT groups, will improve the generalizability of the models and allow for real-world validation.

3. Advanced Deep Learning Architectures:
To enhance classification accuracy, future work will explore more sophisticated deep learning architectures such as Recurrent Neural Networks (RNNs), Bidirectional LSTMs, and Transformer-based models that are better suited for capturing long-range dependencies in sequential data. Moreover, exploring lightweight architectures like MobileNet or quantized CNNs can facilitate deployment in resource-constrained environments such as edge devices and embedded security appliances.

4. Explainability and Model Interpretability:
As AI models are increasingly integrated into security operations, explainability becomes essential. Future work will involve the use of explainable AI techniques—such as SHAP (SHapley Additive exPlanations) and LIME (Local Interpretable Model-agnostic Explanations)—to provide insight into the decision-making process of both traditional and deep learning models. This will help cybersecurity analysts trust and validate the classification outputs, and possibly identify new malicious OpCode patterns.

5. Transfer Learning and Pretrained Embeddings:
Another strategic direction is the application of transfer learning, where models pre-trained on large corpora of assembly or OpCode sequences can be fine-tuned on smaller, task-specific malware datasets. This approach may significantly improve performance in low-data scenarios and accelerate training.

6. Real-Time Detection and Deployment:
The long-term goal is to integrate the developed models into a real-time malware detection pipeline. This would involve optimizing inference time, reducing false positives, and ensuring compatibility with endpoint security solutions. Additionally, deploying the model within a cloud-native or containerized framework would enable scalable and distributed analysis of incoming files in enterprise environments.

In summary, the future roadmap aims to advance this research from a controlled experimental setting toward a practical, enterprise-ready malware detection solution. By leveraging advanced ML/DL models, dynamic threat signals, and explainable AI, this research will contribute to the development of next-generation cybersecurity defenses tailored for APT detection and beyond.

\section{Conclusion and Future Work}
This research presents a comparative study of OpCode-based malware classification using both traditional machine learning algorithms and a deep learning-based CNN. By leveraging static analysis techniques, specifically focusing on OpCode sequences extracted from disassembled malware samples, we investigated the effectiveness of both manually engineered features (n-grams) and automated feature learning. Among the traditional classifiers, the SVM demonstrated the highest accuracy and overall performance, with an F1-score of 64.04\%, outperforming K-Nearest Neighbors and Decision Tree classifiers. The SVM model’s robustness can be attributed to its ability to find optimal hyperplanes in high-dimensional feature spaces, making it particularly suitable for n-gram frequency data. The Voting Classifier ensemble offered modest performance gains but did not surpass SVM alone. The CNN model, while slightly underperforming compared to the SVM in terms of accuracy (62.14\%), offers significant advantages in terms of automation and scalability. By eliminating the need for manual feature engineering, the CNN streamlines the malware classification pipeline and has the potential to uncover deeper, hierarchical patterns in OpCode sequences. However, its performance was likely constrained by dataset size and model capacity, suggesting that deeper architectures or larger training datasets may improve results. This study establishes that OpCode-based static analysis remains a viable approach for malware detection and classification, especially when paired with efficient feature engineering or deep learning architectures. Both traditional and deep learning methods have demonstrated their utility in classifying malware families associated with Advanced Persistent Threats (APTs), offering practical insights for security analysts and automated defense systems. For future work, several avenues can be pursued. First, the exploration of more advanced deep learning architectures, such as Transformers or hybrid CNN-LSTM models, could improve classification performance and better capture long-range dependencies in OpCode sequences. Second, integrating dynamic analysis features (e.g., API calls, system behavior traces) could enhance detection accuracy by combining both static and behavioral perspectives. Third, leveraging transfer learning or pre-trained embeddings for OpCodes may offer improvements in generalization, especially across diverse malware families.
Moreover, incorporating explainable AI (XAI) techniques—such as SHAP or LIME—could improve transparency in decision-making, aiding analysts in understanding which features or OpCode patterns are most influential in classification. Expanding the dataset and including real-world samples from emerging threats would further validate the robustness of the proposed models.

\section*{Additional Information}
\begin{itemize}
    \item \textbf{Implementation Details:} Full code implementations are available in the accompanying files for both traditional machine learning pipelines and the CNN model training process.
    \item \textbf{References:} Key literature including McLaughlin et al. (2017), Bilar (2007), among others, supports the methodologies used.
\end{itemize}

\bibliographystyle{IEEEtran}
\bibliography{ref}
\end{document}